\documentclass[sigconf, preprint]{acmart}
\AtBeginDocument{%
  }

\setcopyright{acmlicensed}
\copyrightyear{2018}
\acmYear{2018}
\acmDOI{XXXXXXX.XXXXXXX}
\acmConference[Conference acronym 'XX]{Make sure to enter the correct
  conference title from your rights confirmation emai}{June 03--05,
  2018}{Woodstock, NY}
\acmISBN{978-1-4503-XXXX-X/18/06}

\begin{document}

\title{SEMANTIC SEE-THROUGH GOGGLES:\\Wearing Linguistic Virtual Reality  in (Artificial) Intelligence}


\author{Goki Muramoto}
\affiliation{%
  \institution{The University of Tokyo}
  \city{Tokyo}
  \country{Japan}
}
\email{goki.muramoto@star.rcast.u-tokyo.ac.jp}

\author{Yuri Yasui}
\affiliation{%
  \institution{The University of British Columbia}
  \city{Vancouver}
  \country{Canada}
}
\email{yyasui@student.ubc.ca}

\author{Hirosuke Asahi}
\affiliation{%
  \institution{The University of Tokyo}
  \city{Tokyo}
  \country{Japan}
}
\email{hirosuke.asahi@star.rcast.u-tokyo.ac.jp}


\renewcommand{\shortauthors}{Muramoto et al.}

\begin{abstract}
When language is utilized as a medium to store and communicate sensory information, there arises a kind of radical virtual reality, namely "the realities that are reduced into the same sentence are virtual/equivalent." In the current era, in which artificial intelligence engages in the linguistic mediation of sensory information, it is imperative to re-examine the various issues pertaining to this potential VR, particularly in relation to bias and (dis)communication.
Semantic See-through Goggles represents an experimental framework for glasses through which the view is fully verbalized and re-depicted into the wearer's view. The participants wear the goggles equipped with a camera and head-mounted display (HMD). In real time, the image captured by the camera is converted by the AI into a single line of text, which is then transformed into an image and presented to the user's eyes. This process enables users to perceive and interact with the real physical world through this redrawn view. We constructed a prototype of these goggles, examined their fundamental characteristics, and then conducted a qualitative analysis of the wearer's experience. This project investigates a methodology for subjectively capturing the situation in which AI serves as a proxy for our perception of the world. At the same time, It also attempts to appropriate some of the energy of today's debate over artificial intelligence for a classical inquiry around the fact that``intelligence can only see the world under meaning''
\end{abstract}


\keywords{Do, Not, Us, This, Code, Put, the, Correct, Terms, for,
  Your, Paper}
\begin{teaserfigure}
   \centering
  \includegraphics[width=\linewidth]{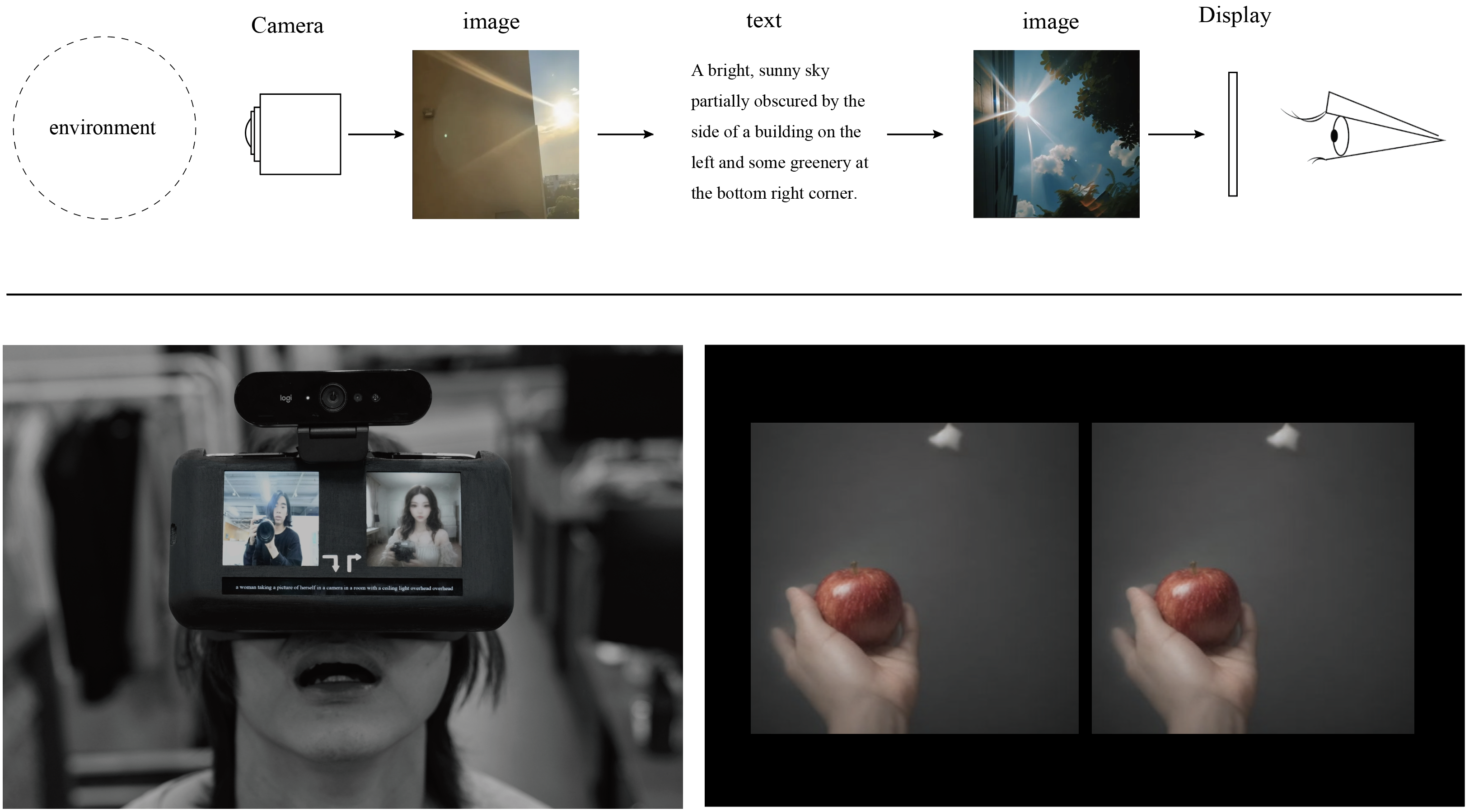}
  \caption{Top: The framework of Semantic See-through Goggles. Bottom-left: The prototype of Semantic See-though Goggles. Bottom-right: the view the wearer is seeing.}
  \label{fig:teaser}
\end{teaserfigure}

\received{20 February 2007}
\received[revised]{12 March 2009}
\received[accepted]{5 June 2009}

\maketitle

\section{Introduction}

Language serves as a medium for sensory information. The utilization of words enables us to store and share the world we experience with others. Since antiquity, humans have employed this medium to transport sensory data across vast distances in both space and time\cite{jowett1892charmides, herder2019treatise, mcluhan2011gutenberg}. Even in the present era, with the advent of technology that directly stores and transmits sensory information through visual and audio media, the continued effectiveness of this medium is evident. However, this medium of language, along with its capacity to store and convey information, also entails the distorted transformation of information. The conversion of sensory information to words entails the selection and reduction of information, whereas the decompression of sensory information from words involves the imposition and supplementation of information. Both processes inevitably entail the assumption and prejudice embedded in the entities and their communities that perform them\cite{bourdieu1991language}. In this process, there is a kind of radical concept of virtual reality, so to speak: "Realities that become the same text when put into words are equivalent/virtual." This goes beyond the issue of language as a medium; it is an inescapable fact that the intellect can only experience the world under a semantic/linguistic umbrella.


Today, artificial intelligence is beginning to participate in this semantic/linguistic mediation process of sensory information. Practical research on the verbalization of sensory information continues to focus on image captioning, which converts images into text\cite{sharma2023comprehensive, https://doi.org/10.48550/arxiv.2201.12086}. In recent year, although somewhat later, the restoration of sensory information from verbal information has also been studied in the context of the automatic generation of images, videos, and music\cite{rombach2022high, li2024comprehensive, luo2023latent}. In the near future, they are said to catch up with human quality\cite{baum2011long}. Even in many cases where explicit verbalization is not involved, the process often includes a latent form of semantic compression or expansion. These processes performed by AI have the potential to be much faster than those carried out by humans, and AI will be a proxy for much of the linguistic mediation of human culture in the future. When this happens, the problems already mentioned with language as a medium for sensory information will become problematic at an accelerated rate. Many of the biases in AI, which numerous researchers have already pointed out,\cite{kim2023mitigating,munoz2023uncovering,bhargava2019exposing,marinucci2023exposing, ntoutsi2020bias} can be understood as manifestations of the problems just described.


Semantic See-through Goggles is an experimental framework for glasses through which the view is once fully verbalized and re-imaged. Participants wear original goggles with a camera and HMD. 
This allows for a subjective understanding of situations in which perceptual information is mediated by AI, or in which perceptual information is mediated by intelligence on the basis of words and meanings, by housing the unit of its processing in goggles through which it views the world.
Imagine there are two little artists inside the goggles: the first is the writer, who describes the scene coming through the camera lens in a straightforward one-line sentence; the second is the painter, who paints a picture based on that text. This writer and painter, or the two adaptors, work at high speed in a different time stream from ours; Artificial Intelligences play the role of these two dwarfs. 
This project is motivated by two attempts. The first is to explore a methodology for subjectively capturing the situation in which AI is a proxy for our perception of the world, and the second is to calmly confront the contemporary debate over AI with the problem that the very fact that "we can only see the world under meaning" already poses.

This paper first reviews the current status of interconversion between images and text as intermedia adaptation by AI and see-through HMDs as goggles to view the environment in front of us. Next, we propose a framework for semantic see-through goggles. Next, we evaluate the implementation of a prototype based on this framework, and confirm and analyze its basic properties as a transformation that meets the framework's stipulations. We then conducted a workshop study in which participants wearing the prototype actually observed and acted in the surrounding space, and conducted a qualitative analysis of their experience. Finally, we discuss the above results, explore the possibilities and limitations of the Semantic See-through framework, and propose the concept of Semantic Virtuality to consider human-AI communication over meaning.

\section{Related works}

\subsection{Adaptation by AI}
The process of transforming content across media is referred to as "adaptation", and is typified by the adaptation of a novel into a film or a film into a novel\cite{holmes1972cross}. 
Insofar as adaptation is an act of arbitrarily separating the "meaning to be preserved" from the "meaning not to be preserved" and changing the medium in which it is expressed, "refraction\cite{lefevere2016translation}" occurs. During this process, the adaptor faithfully tries to move the "meaning of the work." However, the adaptor's cultural background inevitably influences their interpretation of the context of the original work.
Therefore, this task requires a deep understanding of the differences and identities between the two media, and has traditionally been performed by professionals with expertise in adaptation. Recent advancements in multimodal artificial intelligence (AI), capable of translating between images, text, and audio, can be seen as a precursor to AI-driven adaptation. This section focuses on adaptations between images and text, which will play a central role in this project and will be incorporated into its implementation.

\subsubsection{Image captioning}
Image captioning is a task that takes a single image as input and generates a caption sufficient to describe salient details, such as events or human behavior depicted in the image.
In the early days, the task of understanding images by computer was dominated by the approach of generating high-level semantic information(scene, object, or relationship) through low-level visual features(color, texture, and shape). However, with the rise of deep learning in recent years, this approach has shifted. It is now possible to utilize large datasets mapping images to text and directly learn the relationships between images and textual descriptions using neural networks and the Encoder-Decoder model\cite{sharma2023comprehensive}.
At the same time, image captioning has evolved dramatically by constructing many evaluation metrics for generated results, such as BLEU\cite{reiter2018structured} and CIDEr\cite{vedantam2015cider}, to compare results with human-generated ground truth.
Although the quality of caption generation has yet to be at the human level, many applications have already been proposed. For example, image captioning is used to generate optimal titles for product images\cite{Harzig2018Multimodal} and to generate captions from images embedded in news articles\cite{Tran2020Transform}.
In addition, the process speed of image captioning also opens up applications. For example, image captioning is used in caption generation methods for the visually and hearing impaired\cite{Makav2019Smartphone-based} and in scene identification systems for more immediate human understanding of visual information\cite{Hsieh2019Implementing}.
As image captioning, approches near-human levels of quality and speed, it is gradually beginning to naturally take over the human task of describing the world.

\subsubsection{Image generation}
Image generation is a task that takes natural language as input and outputs a single image of an event, person, or other behavior described in the text.
In the early days, the task of generating images by computer was dominated by the approach of directly commanding a specific act of drawing, as symbolized by drawing machines with built-in pen and ink\cite{noll1994beginnings}. However, as commands have become more abstract in recent years, it is now possible to instruct a computer to generate an image by describing the object to be depicted using natural language.
The advent of deep learning is said to have removed major limitations in both quality and speed of image generation. Among the various models, such as GANs and VAEs, the emergence of Stable Diffusion\cite{rombach2022high}, which employs a denoising diffusion probabilistic model to transform random noise into high-quality images, makes a significant milestone in generative model technology\cite{li2024comprehensive}. Consequently, various applications are being considered.
For example, text-to-image generation improves efficiency in the marketing field\cite{haleem2022artificial} and assists the early stages of the architectural design\cite{paananen2023using}.
Furthermore, the faster image generation is also expected to find applications in fields such as animation, which requires a rapid design process\cite{ding2022cogview2}.
As with image captioning, image generation is increasingly approaching human quality and speed and taking on human creativity.

\subsubsection{Bias in AI}
Whether in image captioning or image generation, the inference of the model is known to be biased. In image captioning, it has been observed that generated captions can be biased because the model relies more on contextual evidence from the dataset than on visual features\cite{kim2023mitigating}. In image generation, the evident imbalance in face generation performance across different social groups presents a significant issue\cite{munoz2023uncovering}. 
However, these bias problems are universal, given that Intelligence (I in AI) is basically based on some artificial(A in AI) data. Even in the frequently used COCO caption dataset, gender bias and racial bias have been noted\cite{bhargava2019exposing}. The fact that a dataset is artificial implies that AI learning is inherently based on selective sampling by a specific human group, where certain aspects of the world are included while others are omitted. As several studies have pointed out, AI can inherit human biases\cite{marinucci2023exposing, ntoutsi2020bias}. And in this process, these biases are unintentionally embedded. Despite AI's increasing presence in society, these issues remain unresolved, both practically and theoretically. AI biases can perpetuate and reinforce stereotypes, leaving certain groups behind in their economic, political, and social lives. For example, a study on the Correctional Offender Management Profiling for Alternative Sanctions (COMPAS) software, used in the United States to determine whether to release an offender, found that the system was biased against African Americans. African American offenders were more likely to be assigned a higher risk score than white offenders with the same profile\cite{dressel2018accuracy}.

In recent years, there have been increasing efforts to address issues related to bias in AI. Since fairness is often context-dependent, fully automated methods are considered limited in their effectiveness \cite{wachter2021fairness}. As a result, human-in-the-loop (HITL) approaches, which incorporate human feedback, have gained significant attention. HITL is expected to play a crucial role in mitigating AI bias and ensuring decision-making transparency, as it introduces human intervention at various stages of AI systems, from data training to inference.
As far as the data training phase is concerned, workflows have been proposed to identify biases in image datasets through crowdsourcing\cite{Hu2020Crowdsourcing}.
There are also approaches to focus on the algorithmic properties of the model, such as methods for discovering unknown bias attributes through the latent space of generative models \cite{Li2021Discover}, and methods for visualizing the attention mechanisms of transformer models \cite{Vig2019Visualizing}. A particularly intuitive approach to uncovering bias involves directly comparing AI inputs and outputs, such as the source image and its generated caption in image captioning\cite{Bakr2023ImageCaptioner2:, Kasai2021Transparent}or the prompt and generated image in image generation\cite{Wu2023Human, Otani2023Toward}.

The Semantic See-Through Goggle proposed in this study can be positioned as an extension of the HITL method, which directly compares AI input and output. Specifically, the goggle allows the input image and the generated image to be viewed side by side, with both capable of being converted into sentences carrying the same meaning. Since this method integrates the two tasks of image captioning and image generation, the final output image reflects the accumulated biases of both models. However, the ability to compare images on the same semantic layer (i.e., image-to-image comparison) and the potential to engage with real-world visual experiences through the goggle may offer advantages over existing methods in addressing AI bias.

\begin{figure}[tb]
 \centering 
 \includegraphics[width=\columnwidth]{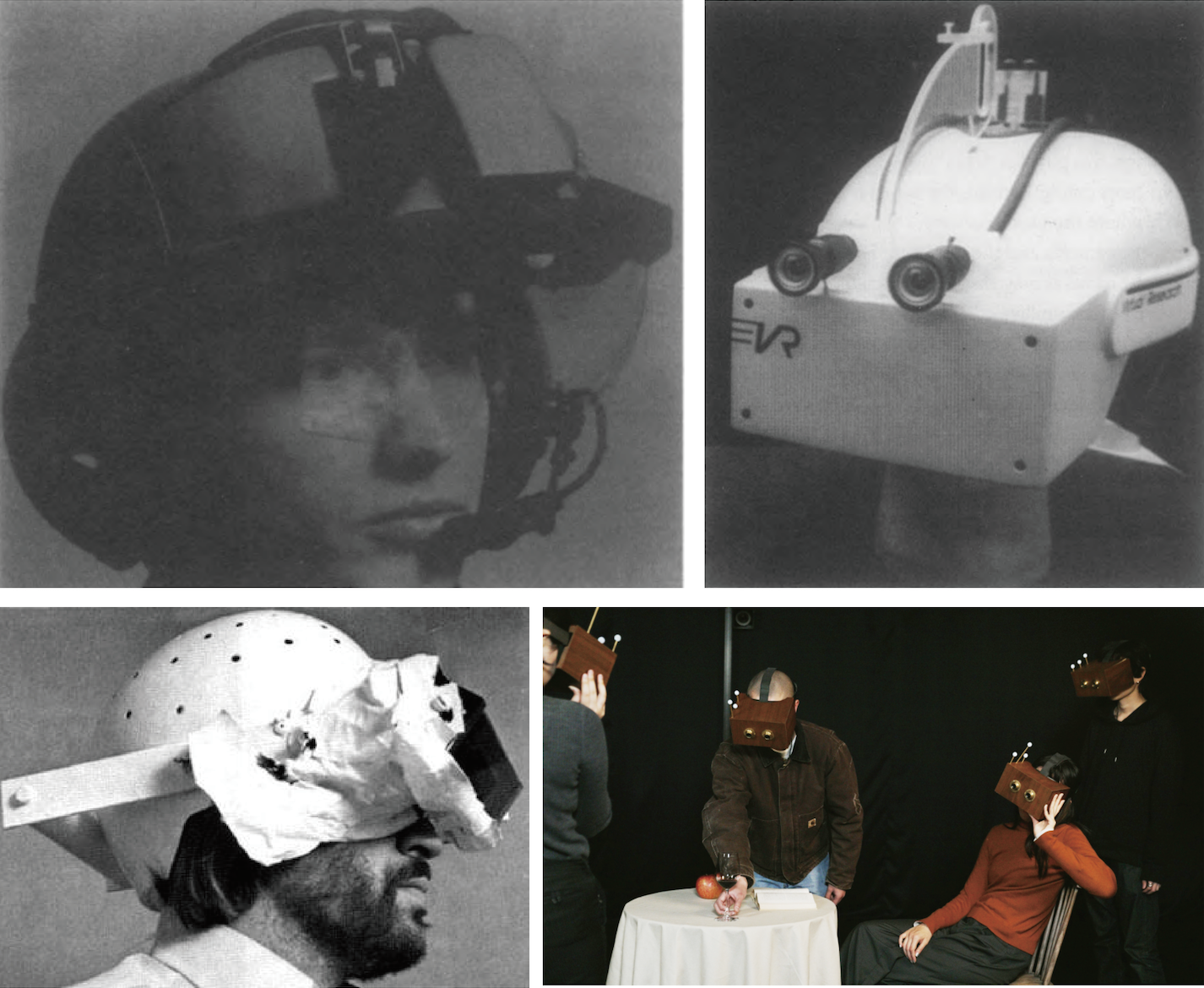}
 \caption{See-through HMD is a goggle to see what is in front of you. Top-Left: an optical see-through HMD made by Huges Electronics. Top-right: a video see-through HMD by Courtesy Jannick Rolland, Frank Biocca, and UNC Chapel Hill Dept of Computer Science (Photo by Alex Tremi). Bottom-left: Inverted glasses experiments. Bottom-right: Lived Montage by Goki Muramoto (Photo: Kai Fukubayashi).}
 \label{fig:seethrough}
\end{figure}

\subsection{See-Through HMD}


See-through is the name used for the system ``through'' which one ``see''s the environment(Fig.\ref{fig:seethrough}). 
There are usually two types of see-through: optical see-through, which uses light coming directly from in front of the eyes, and video see-through, which captures the light once with a camera.\cite{azuma1997survey}. 
By making diverse edits (adding, deleting, or modifying information) to these system processes, the various image editing technologies are transformed into perceptual augmentation technologies through our views, thereby extending reality\cite{bajura1992merging}.

See-through, however, can also be interpreted more expansively and radically. 
If we rethink the question of "what is being seen," it is not always necessary to use the incoming image directly as a base, and more radical replacement is possible.
In the inverting glasses experiment, goggles with upside-down or left-right reversed will be provided.\cite{stratton1896some}. 
However, the user gradually comes to understand and act upon the environment through the goggles, and the experience of seeing is preserved.
The Lived Montage project involves the experience of looking at the view of others who are looking at the same thing, and participants can understand the physical space they belong to and act on it through the perspectives of others\cite{Goki2020}. These are also see-through goggles, considering that they are goggles for seeing one's own environment in front of them, by the rule of maintaining the object of looking. Similarly, this study tentatively answered the question of what is seen with "meaning" and expressed it in "one sentence" of text.

\section{Conceptual framework}
Semantic See-through Goggles are defined as a kind of video see-through performed by the following procedure.

\begin{enumerate}

\item \textit{View capturing}: 
The image is captured from the camera near the eyes.

\item \textit{Adaptation1 (Image to Text)}: 
The image is translated into one sentence.

\item \textit{Adaptation2 (Text to image)}: 
The sentence is translated into one image.

\item \textit{View presentation}: 
The image is presented to the eyes

\end{enumerate}

The entity that makes the two adaptations is not included in the abstract definition, but since the entire procedure needs to be done in real-time and at high speed, Artificial Intelligence will be responsible for these adaptations in reality.



\section{Implementation}
This section presents examples of Semantic See-through Google implementations, divided into their system design and model selection.
\begin{figure}[h]
 \centering 
 \centering 
 \includegraphics[width=\columnwidth]{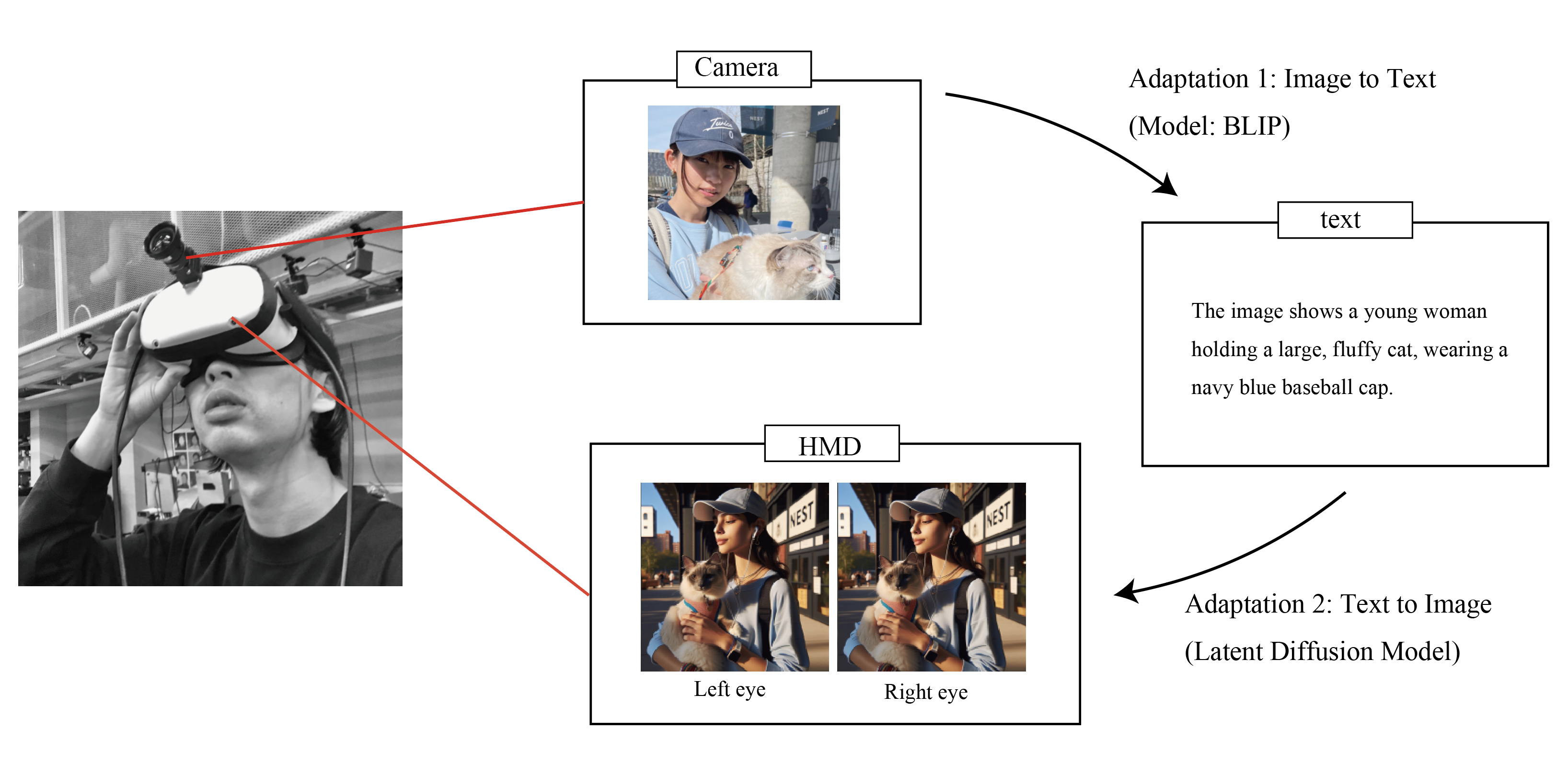}
 \caption{The sketch on the implementation of Semantic see-through Goggles. An HMD with a camera is provided, and on the way the landscape image arrives from the camera to the HMD, it is converted once into a single line of text and re-imaged, by the two AIs.}
 \label{fig:system}
\end{figure}

\subsection{System design}
The semantic see-through goggles consist of an original head-mounted display (HMD) with inner and outer displays, a camera capable of video capture, and a computer for software processing. (Fig.\ref{fig:system},\ref{fig:hmd}) The transfer of information from the camera to the computer and from the computer to the internal and external displays was done by wire. According to the proposed framework, the system performs the following processes. First, real-time camera images are captured and stored in a buffer on the computer. Then, the buffered images go through two AI-powered transformations: the image is converted to text, and the text is then used to generate a new image, all according to the framework. The generated image is displayed in the same way on the inner left-eye and right-eye displays of the HMD, respectively, and the viewer sees only the adapted image, not the original image. On the other hand, the outer display of the HMD shows the original image captured by the camera, the text converted from the image, and the generated image without any distortion in the combination of the three types of content.Thus, participants viewing the goggle wearer from the outside can perform various provocative actions while watching the adaptation process displayed on the HMD's outer display.

\subsection{Algorithm}
Our system uses a published high-speed AI model to minimize delays in the two transformations. For image captioning, we use BLIP\cite{https://doi.org/10.48550/arxiv.2201.12086}, a model pre-trained on a large boosted trap dataset, to caption the scene of a single image in a single sentence. The model is pre-trained on both visual data, offering high-quality visual representation and linguistic data, providing strong language generation and zero-shot transfer ability. We limited the length of the sentences in image captioning to between 20 and 40 words.
For image generation, we use LCMs\cite{luo2023latent}, a model that streamlines the inference process in the Latent Diffusion Model (LDM), to generate images from single sentences. This model enables rapid inference for pre-trained LDMs with a minimal number of steps. We adjusted the parameters so that the inference process in image generation runs 4 steps.
In addition, both input and output images were standardized to 640 x 640 pixels. As a result, the time required for both processes is limited to about one second, enabling the system to follow changes in see-through scenery almost in real-time while maintaining image quality. 




\section{Preliminary Study}
This section provides a preliminary quick survey of the characteristics of transformation in a semantic see-through framework. Specifically, we investigated the loss and retention that occur during the process of reconverting an original image into an image via text based on quantitative measures and human questionnaires.

\subsection{Image Database and model parameters}
The database used is the freely available DET test dataset from ILSVRC 2017. ILSVRC was a large-scale object detection and image classification challenge that ran from 2012 to 2017\cite{russakovsky2015imagenet}. The images were center-cropped and resized to 256x256 pixels to match desired output size after image conversion. The images were randomly selected, resulting in a diverse mix of subjects, including people, animals, and landscapes.
The parameters for image captioning and image generation are aligned with those used in real-time applications. Therefore, the length of the captioned text is kept within 20-50 characters, and the inference step for image generation is limited to 4 steps. The output image is adjusted to 256 x 256 pixels to maintain consistency with the input image dimensions.

\subsection{Consistency of linguistic semantics}
In Semantic See-through Google, images are converted into sentences and then back into images. To maintain consistency, the input and output images should have the same meaning when turned into sentences. This consistency is also an important and fundamental property when considering Semantic virtual reality, which states that ‘a reality with the same linguistic meaning is a reality’. In the present implementation, it is necessary to check to what extent this property is realised, as image captioning and image generation are not completely symmetric processes with the same model.

\paragraph{Method}
In addition to captioning the pre-transformed images in the Semantic See-through Goggle, the post-transformed images were also captioned using the same method. To quantify what is preserved and lost linguistically, we calculate sentence similarity using several indices. These scores are then compared to those from random sentence combinations.

Phrases and words that are continuously repeated in the sentences were eliminated as a preprocessing step, leaving only one since they are noise in the comparison. We used four metrics: (i) TF-IDF, (ii) Word Mover's Distance (WMD), (iii) Universal Sentence Encoder (USE), and (iv) Sentence-BERT (SBERT).
TF-IDF scores the importance of words based on their frequency of occurrence and inverse document frequency, assigning higher scores to rarer words across the documents. WMD evaluates semantic similarity between sentences by calculating distances in a word embedding space\cite{kusner2015word}. USE utilizes a transformer architecture to embed entire sentences, capturing meanings more effectively than word-level embeddings\cite{cer2018universal}. SBERT builds on BERT's architecture to encode sentences individually, capturing their detailed meanings\cite{reimers2019sentence}. For word-based comparison metrics like TF-IDF and WMD, preprocessing involves stop-word removal and tokenization. Cosine similarity is used to measure similarities in TF-IDF, USE, and SBERT, assessing the angles between embedded vectors. For WMD, similarity calculation is performed by taking the reciprocal of the distance; specifically, the similarity is 1 when the distance is 0, and it approaches 0 as the distance increases. For comparison, 2,500 pairs of sentences captioned from images before and after conversion were used.

\paragraph{Result}
\begin{figure}[t]
 \begin{center}
  \includegraphics*[width=\columnwidth]{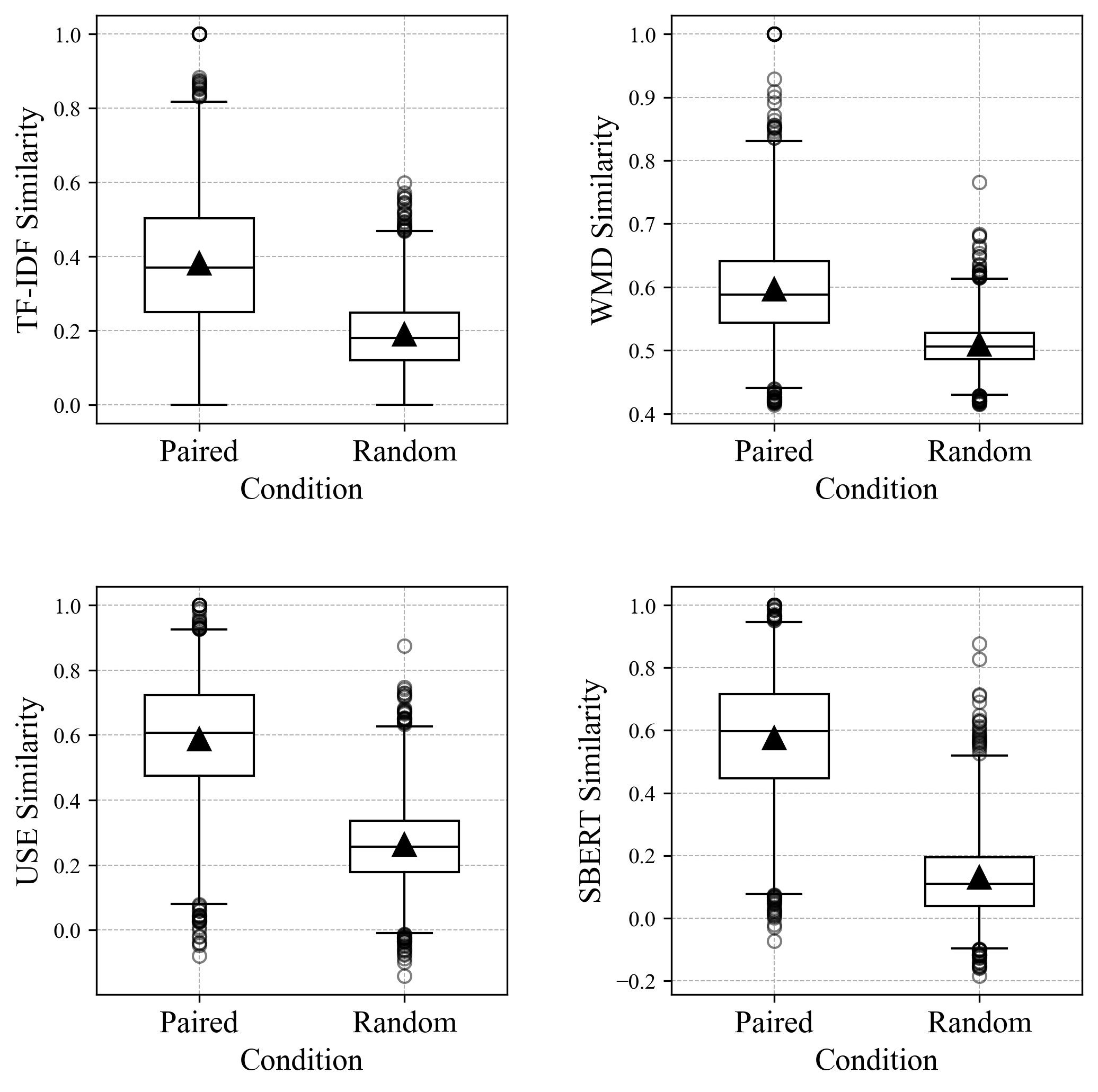}
 \end{center}
 \caption{Comparison of linguistic similarities using different similarity indices.The top left figure shows the distribution for each condition in TF-IDF, the top right shows WMD, the bottom left shows USE, and the bottom right shows SBERT. The vertical axis is not aligned in order to see the distribution. Outliers are defined as values in the 99th percentile.}
 \label{fig:text_similarity}
\end{figure}

The distribution of the similarity of the four similarity indices, TF-IDF, WMD, USE, and SBERT, in the Paired Condition (Correspondence Condition) and Random Condition (Random Condition) is shown in Figure 4\ref{fig:text_similarity}.
Correspondence t-tests were conducted for each indicator, and the results are shown in Table 1\ref{tab:table1}. Scores for all indicators tended to be higher in the Paired Condition than in the Random Condition, and statistically significant differences were found (p < 0.001). The effect size (Cohen's d\cite{cohen2013statistical}) was also large, with Cohen's d greater than 0.8. These results suggest that the image transformation algorithm retains a certain degree of the original image features that each indicator captures in the transformed image.
Additionally, a comparison of absolute similarity values across the measures showed that TF-IDF had a low average similarity of 0.38 in the Paired Condition, whereas WMD, USE, and SBERT exhibited moderate similarities of 0.60, 0.58, and 0.57, respectively. The higher similarity observed for WMD compared to TF-IDF suggests that, while the algorithm does not maintain lexical-level completeness as captured by TF-IDF, it is capable of preserving the semantic similarity between words, as demonstrated by WMD. Furthermore, the moderate similarity scores from USE and SBERT indicate that the transformation algorithm retains a certain amount of higher-order semantic information embedded in the sentences. Notably, the effect size increased progressively from TF-IDF to SBERT, with values of 1.01, 1.09, 1.59, and 1.93, as the indices became more semantic. This trend suggests that the transformation algorithm is more effective at preserving higher-order semantic information than lexical-level meaning.


\begin{table}[h]
\centering
\caption{Statistics for the linguistic similarities}
\label{tab:table1}
\resizebox{\columnwidth}{!}{%
\begin{tabular}{|c||c|c|c|c|c|c|c|}
\hline
\textbf{Indices} & \textbf{Ave\textsubscript{P}} & \textbf{Std\textsubscript{P}} & \textbf{Ave\textsubscript{R}} & \textbf{Std\textsubscript{R}} & \textbf{t value} & \textbf{p value} & \textbf{Cohen's d} \\ \hline
TF-IDF & 0.38 & 0.18 & 0.19 & 0.10 & 51.75 & \textless 0.001 & 1.01 \\ \hline
WMD & 0.60 & 0.08 & 0.51 & 0.04 & 55.44 & \textless 0.001 & 1.09 \\ \hline
USE & 0.58 & 0.19 & 0.26 & 0.13 & 81.21 & \textless 0.001 & 1.59 \\ \hline
SBERT & 0.57 & 0.20 & 0.13 & 0.13 & 98.65 & \textless 0.001 & 1.93 \\ \hline
\end{tabular}%
}
\end{table}

\subsection{Consistency of visual semantics}
In Semantic See-through Goggle, images are transformed back into other images through sentences. As a result, the input and output images should share the linguistic meaning of a single sentence and visually represent their common features. This section examines the characteristics of the system's process when understood as an image transformation.

\paragraph{Method}
To quantitatively assess what is preserved and lost during image transformation, a similarity calculation between the input and output images was performed using several indices. The scores were compared with those from random combinations of images. We used four metrics: (i) Histogram Intersection (HI), (ii) Scale-Invariant Feature Transform (SIFT), (iii) AlexNet-based Learning Perceptual Image Patch Similarity (LPIPS\textsubscript{A}), and (iv) Transformer-based Learning Perceptual Image Patch Similarity (LPIPS\textsubscript{T}). HI is a method used to assess the visual similarity between color histograms of different images by calculating the overlapping areas of their histograms. In this case, the calculation is performed in the CIELAB color space, a color model designed to reflect human visual perception\cite{Lee2005EvaluationOI}. SIFT detects local image features that are invariant to scale and rotation, generating a feature vector. 
LPIPS (Learned Perceptual Image Patch Similarity) is a method that employs pre-trained deep neural networks to extract features and is believed to offer an evaluation of images that closely aligns with human subjective perceptiont\cite{zhang2018unreasonable}. In this study, we utilized both AlexNet and Transformer models for LPIPS. It is important to note that, unlike other indices, a lower LPIPS score indicates higher similarity between images. For each comparison, 2,500 pairs of input and output images were used.

\begin{figure}[tb]
 \begin{center}
  \includegraphics*[width=\columnwidth]{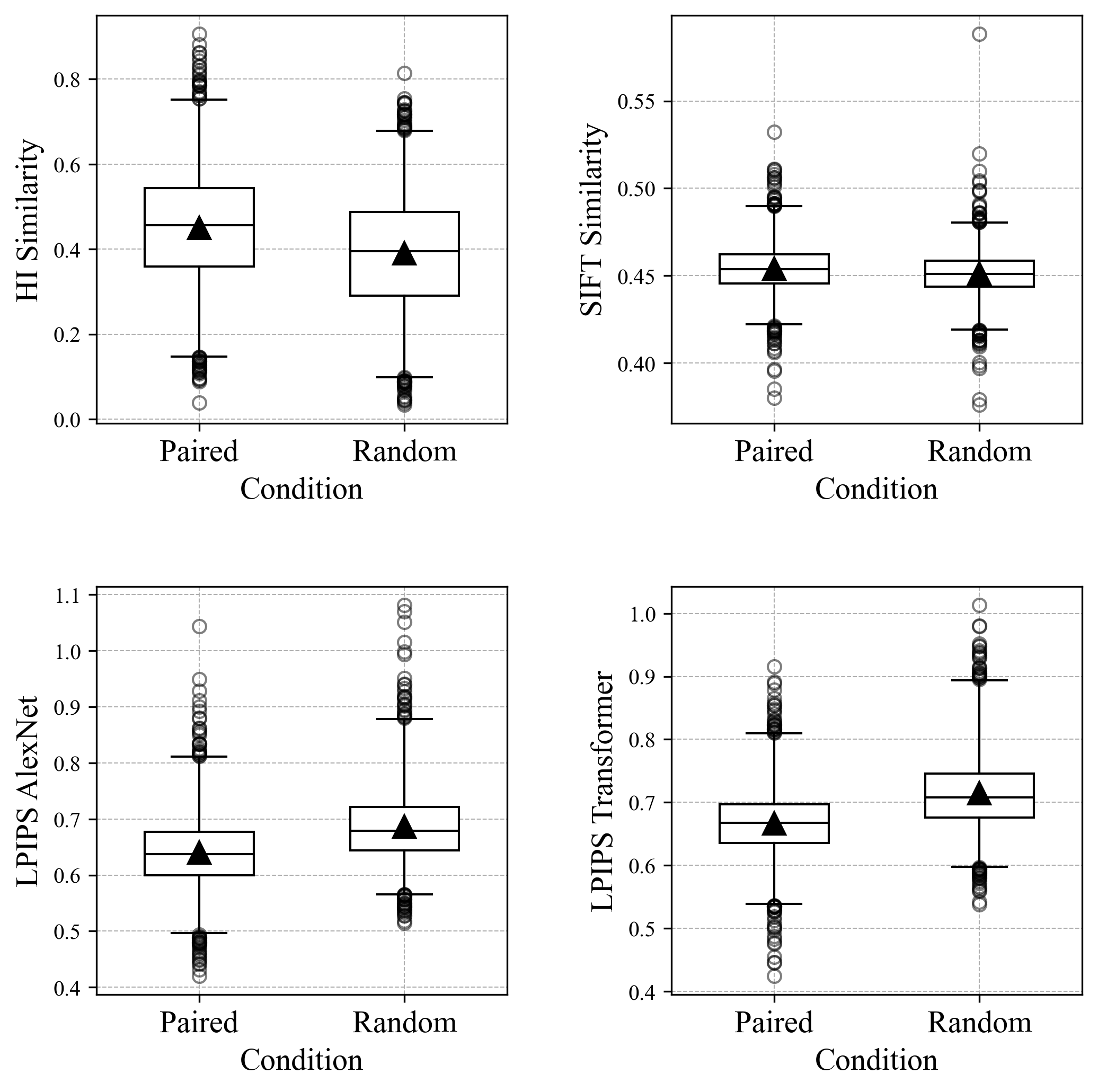}
 \end{center}
 \caption{Comparison of visual similarities using different similarity indices. The top left figure shows the distribution for each condition in HI similarity, the top right shows SIFT similarity, the bottom left shows LPIPS\textsubscript{A}, and the bottom right shows LPIPS\textsubscript{T}. The vertical axis is not aligned in order to see the distribution. Outliers are defined as values in the 99th percentile.}
 \label{fig:image_similarity}
\end{figure}

\paragraph{Result}
The distribution of the similarity of four similarity indices, HI Similarity, SIFT Similarity, LPIPS\textsubscript{A}, and LPIPS\textsubscript{T}, in the Paired Condition (Correspondence Condition) and Random Condition (Random Condition) is shown in Figure 5\ref{fig:image_similarity}.
The corresponding t-test results for each index are provided in Table 2\ref{tab:table2}, where statistically significant differences were observed for all indices (p < 0.001).
For HI Similarity, the analysis yielded a Cohen's d of 0.39, with a mean similarity of 0.45 in the Paired Condition, indicating a moderate effect size. This suggests that the transformation preserves color distribution features to a certain extent.
In contrast, SIFT Similarity exhibited a similar mean similarity, but the difference from the Random Condition was minimal, with a low Cohen's d of 0.19. This indicates that many local features, such as edges and corners, were lost during the transformation.
In contrast, the analysis of LPIPS\textsubscript{A} and LPIPS\textsubscript{T} showed relatively high average scores in the Paired Condition, at 0.64 and 0.67, respectively. Cohen's d values were also large, at -0.79 and -0.75, respectively.
Since higher scores in the LPIPS index indicate lower similarity, these absolute scores confirm that the conversion process does not effectively preserve visual and higher-order semantic information as measured by LPIPS. Nonetheless, the magnitude of the effect sizes suggests that the transformation algorithm has a statistically significant impact on preserving some aspects of this information.
Based on these results, the image transformation in Semantic See-through Goggles alters local features, while color information tends to be preserved. Additionally, the transformation was also found to have a consistent effect on visual and higher-order semantic information.

\begin{table}[h]
\centering
\caption{Statistics for the visual similarities}
\label{tab:table2}
\resizebox{\columnwidth}{!}{%
\begin{tabular}{|c||c|c|c|c|c|c|c|}
\hline
\textbf{Indices} & \textbf{Ave\textsubscript{P}} & \textbf{Std\textsubscript{P}} & \textbf{Ave\textsubscript{R}} & \textbf{Std\textsubscript{R}} & \textbf{t value} & \textbf{p value} & \textbf{Cohen's d} \\ \hline
HI Similarity      & 0.45 & 0.13 & 0.39 & 0.13 & 19.40  & \textless 0.001  & 0.39 \\ \hline
SIFT Similarity    & 0.45 & 0.01 & 0.45 & 0.01 & 9.65   & \textless 0.001  & 0.19 \\ \hline
LPIPS AlexNet      & 0.64 & 0.06 & 0.69 & 0.06 & -39.43 & \textless 0.001  & -0.79 \\ \hline
LPIPS Transformer  & 0.67 & 0.05 & 0.71 & 0.06 & -37.56 & \textless 0.001  & -0.75 \\ \hline
\end{tabular}%
}
\end{table}

\subsection{What is being preserved/lost?}
In semantic-see-through-goggle, the input image is transformed into the output image via a textualization process, so essentially, only the content of the text is preserved, with the rest of the content either lost or replaced by something else. The content of the text is a general description of the scene according to the image-captioning model, but it is not known what properties of the image this content refers to for humans. Therefore, we conducted a quick investigation of what is preserved and what is lost in input-output images for humans.

\paragraph{Method}
Subjects were recruited using an online platform, and a web-based questionnaire was administered. Gender, age, and nationality were collected as basic demographic information for the subjects. Participants were asked to observe 200 pairs of input and output images for 10 minutes and then answer the following three questions.
Note that all participants were unaware of our algorithm, i.e., they answered the questions by looking only at the input and output of the images.
A total of 24 respondents participated in the study, with ages ranging from 22 to 25. Of the respondents, 20 were male and 4 were female. Regarding nationality, 23 were Japanese, and 1 was Chinese.


\begin{table}[h]
\centering
\caption{Item Statements}
\begin{tabular}{|c|p{7cm}|}
\hline
\textbf{Item} & \textbf{Statement} \\ \hline
Q1 & What do you consider to be preserved in the image before and after conversion? \\ \hline
Q2 & What do you consider to be lost in the image before and after conversion? \\ \hline
Q3 & Describe this image conversion freely. \\ \hline
\end{tabular}
\end{table}

\paragraph{Result}
The following elements were extracted from the answers. Although these are important results related to the question, “What can the linguistic mediation of images (by AI) carry?” we limit ourselves to enumeration in this paper and leave detailed analysis for future work. For Q3, a variety of comments were collected on the impressions of the conversion, but their qualitative analysis is also an issue for future work and is omitted from this paper.

\paragraph{Q1}
Situation / Category / Overall theme / Image theme / Subject category / Approximate genre / Main message / Concept / Abstract conceptual information / Main motif / Center-focused subject / Main image / Attention target / Subject of focus / Subject’s objective attributes / Subject characteristics / Feature of the subject / Main subject / Color scheme / Overall color tone / Color tones / Colors / Appearance of colors / Elements in the image / Objects included / Present elements / Main elements / Material / Object information / Object's category / Biological entities / Presence of living beings / Type of beings / Action / Composition / Layout / Pose of people or animals / Situation / Setting / Scene / General content / Number of objects / Approximate number of objects / Distance from the subject / Spatial relationship with the object / Gender / Gender of entities depicted / Location / Place / Image / Impressions conveyed by the image.

\paragraph{Q2}
Details / Margins / Realism / Photo-likeness / Accuracy / Fine details / Clarity / Background / Text / Numerical information / Identity / Overall saturation / Detailed characteristics of the subject / Main subject and background / Specificity / Actions / Movement / Dynamism / Color tone / Saturation / Composition / Image focus / Facial expressions / Race / Specific features of the subject / General theme / Low-importance conceptual information / Concrete information with specificity / Realism / Sense of reality / Style / Time concept / Sequence / Subject's personality / Photographer's intention / Position / Posture / Meta-information about the photo / Overall tone / Human expressions / Presence of reflections or additional elements in the image.


\section{Workshop study}

\begin{figure*}[tb]
 \begin{center}
  \includegraphics*[width=\textwidth]{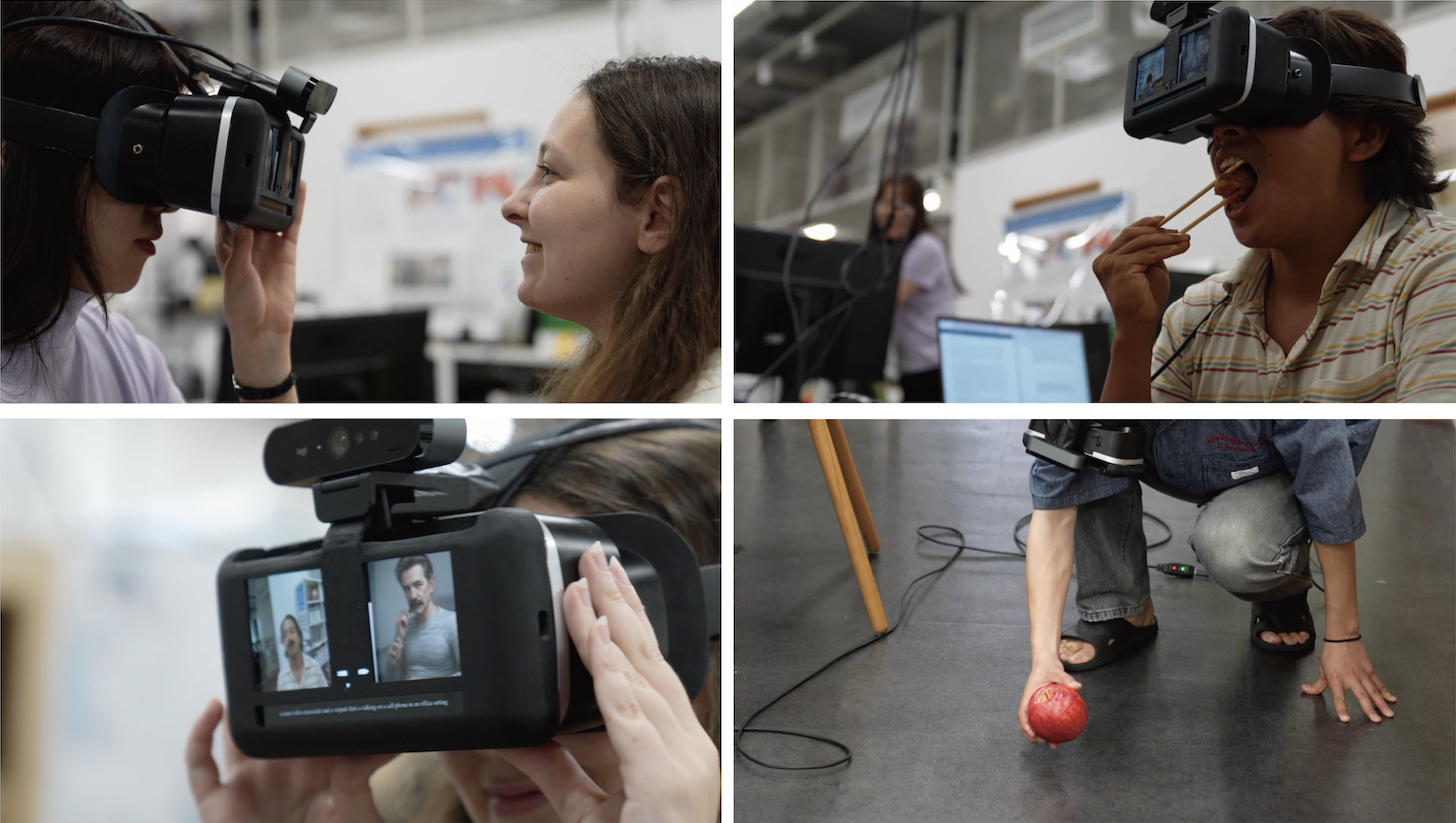}
 \end{center}
 \caption{The workshop view. Participants wear Semantic See-through Goggles and observe, walk, and interact with the environment involving various objects and people. Participants who are not wearing the goggles can see what the wearer is looking at and what text is being mediated by a display embedded in the front side of the goggles.}
 \label{fig:workshop}
\end{figure*}

This section describes a workshop study based on interviews and behavioral observations about the experience of viewing physical reality through Semantic See-through Googles(Fig.\ref{fig:workshop}).
We conducted this workshop study with the same participants as the preliminary experiment (6 males and 3 females; aged 23 - 35, mean age 25 years old, $SD = 3.77$), who agreed to take part in this research after being informed about the study and the system before the demonstration experience.
All participants did not have a physical disability.

\subsection{Workshop Design}
Participants wore the goggles implemented in the above description and interacted in an environment filled with various objects and people. Each participant experienced the system for at least 10 minutes and could continue for up to 30 minutes until they were satisfied. 
The feedback on the system was interviewed via face-to-face and collected from participants by the first author after they completed the experience. During workshop, the authors were observing their behavior in person and on video. No participants declined to participate in the study or the interviews.
This study was conducted with the approval of the ethics committee of the authors' affiliated institution. 

\subsection{Result and Discussion}
In this section, the results of the behavioral observations and the interviews are described and discussed.

\subsubsection{Symboric behaviors}

In the following, the four symbolic events are hilighted(Fig.\ref{fig:workshop}).


\textbf{Watching people}
Semantic See-through Goggles interpret a person in sentences as concise as "a man with black hair standing." It can tell how many people of each gender and what clothing they wear in front of you You also notice that someone new is approaching, and if the person in front of you is wearing glasses, they will appear with glasses in your view. However, the human, in your view, is extremely stereotypical, and that stereotyping depends on the data set the model is using. In this implementation, in most cases, the man is a muscular white male, and the female is a slender white female. However, the depiction of individuals within the scene can vary significantly, as the system's interpretation is biased.  For instance, a slender Asian man with long hair might alternately appear as a white man or a white woman in the view of the goggles wearer.

\textbf{Picking up something (and eat it)}
Although much of the spatial information is lost once converted to text, participants can still engage and interact with the environment. For example, an apple comes into view when you turn around and look down. While you're unsure of its exact position, you move your face closer to it as you reach out. If the apple disappears, you slightly adjust your face in the opposite direction of your previous movement, and the apple reappears. Then, you can pick it up and eat it. It means that the scenery in front of you appears as something that your body works on through it (and even as something that you can eat and make a part of yourself).

\textbf{Looking up}
The text that Semantic See-through Goggles passes through does not contain direct information about perspective. However, when the user looks up, expressions such as "wall and ceiling" and "ceiling" appear in the text, and the image in the field of view gradually changes upward, as if the system is sensing the direction of the head. This experience brings about a strange sensation of one's own spatial movement being symbolized and also allows the goggle wearer to appreciate how camera work is realized in linguistic information such as novels and daily conversations.

\textbf{External initiatives}
The images in the goggle wearer's view and the adaption process are displayed outside the goggles so that they can be seen by the surrounding workshop participants, which prompts them to act. One of the most prominent behaviors is intentionally bringing oneself and the view closer to the stereotype so that it is kept to the output as meaning. For example, if a goggle wearer is looking at his or her own face, the person will actively smile, get angry, or make facial expressions clear. Some participants also dared to give the goggles a different interpretation from reality. For example, they intentionally put on a mustache with black tape to make people estimate their age higher or make stereotypical gestures of the gender they want to appear as.

\subsubsection{Analysis of comments}
The data was developed into three themes, and the experiences and findings were summarized: 1) First, the proposed system is established as a see-through goggle through which one can observe and act upon one's surroundings. Then, 2) we will show how, through this experience, the participants subjectively understood the characteristics and problems of AI adaptations, and 3) how they saw part of it as a general problem of mediating sensory information through language, beyond AI.

\vskip\baselineskip

\textbf{See-through Goggles}

Participants were able to observe their environment through Semantic See-through Goggles. Although much of the information was rewritten in the implementation, the verbalized information allowed them to vaguely understand their surroundings and spatial structure. They felt that the goggles were "see-through," especially when they were consistent with their own "looking around," such as when they looked up and saw the ceiling or when they looked to the side and saw things moving backward.

\begin{quote}
P4: \textit{``
It gives a rough idea of the environment, such as a person in the picture or a bookshelf in the picture. It is interesting to note that there is no direct sensing of directional information, but if I look up, I can see the ceiling.
''}
\end{quote}



\begin{quote}
P8: \textit{``
When wearing the goggles, the experience became what could be called “first-person.
''}
\end{quote}

\begin{quote}
P9: \textit{``
Even though the image changes rapidly, the sense that I am looking from this point of view, that it is a first-person perspective, is maintained.
''}
\end{quote}

Some participants emphasized the first-person and real nature of this experience by comparing it to the experience of looking at another participant's head monitor (which displays what the person sees) when they are not wearing it.

\begin{quote}
P4: \textit{``
It was a different experience when I wore the headset than when I saw the image on the monitor of the other person's headset. When I wore the goggles, I could feel that this was a copy of information from the real world, and I felt like I was focusing more on the content and getting information from the environment.
''}
\end{quote}

Participants were not only able to further understand their surroundings, but were also able to reach out, walk, and perform actions based on them. However, directions, etc., did not appear precisely because they were via rough linguistic information such as “right,” “left,” and “next to something,” etc. The participants felt frustrated by this, but they were able to connect with the AI-generated reality. The participants felt frustrated by this, but they connected the images created by the AI with reality.

\begin{quote}
P5: \textit{``
It is impossible to walk in the same way as usual, but the minimum requirement of communicating and working with things and people is guaranteed through this process.
''}
\end{quote}

\begin{quote}
P7: \textit{``
The ball is right in front of me as my line of sight, but I felt frustrated there because it is not definitely reflected there as an image, but the frustration was strange. Somehow it must be in that direction, and I relied on the sensation of my hand for the rest.
''}
\end{quote}

\begin{quote}

P9: \textit{``
As I picked up the ball, I somehow felt the direction and reached for it. The moment I found it, the distance between the camera position and my ball was all connected.
''}
\end{quote}

Furthermore, some participants reported that as they focused on the action, they became less concerned with the transformation of reality and the dizzying changes in the background caused by the processing.

\begin{quote}

P1: \textit{``
Sometimes, when I try to move based on the goggles' view, I don't mind the surprisingly detailed colors and shapes.
''}
\end{quote}

\begin{quote}

P7: \textit{``
The goggles correctly display the elements that you would pay attention to if you were there, so once you get used to them and narrow down what you would normally pay attention to, it becomes easier to move around.
''}
\end{quote}

The experience of using images transformed through verbalization for understanding the real world and for one's own actions often led to the confusing question, “What is it that I am seeing now?

\begin{quote}

P6: \textit{``
I see a book here. I see the concept of a book, but it is not tied to a specific book. It's an image, so it's not a concept?
''}
\end{quote}

\begin{quote}

P1: \textit{``
What are we looking at here? It clearly means reality, but it is not reality. It is neither concrete nor abstract...
''}
\end{quote}

\vskip\baselineskip

\textbf{Subjective understanding of AI}

Many participants initially linked their discomfort compared to their everyday visual experiences to the technology of AI and the biases of its data and processing. The most strongly felt features were that what is important (i.e., the value of information) and how events are isolated are determined based on bias, and that when abstracted information is restored to concrete images based on bias, it includes beautification and averaging. Many participants also experienced fear and disgust as they experienced this bias in the first-person perspective.

\begin{quote}

P3: \textit{``
Is this the view seen from the AI?
''}
\end{quote}

\begin{quote}

P1: \textit{``
I felt that I was subconsciously aware of algorithmic bias. By seeing it as my view, the bias I knew as knowledge manifested itself differently.
''}
\end{quote}


\begin{quote}

P5: \textit{``
Various things we see tend to be beautified. For example, my own face and body were beautified.
''}
\end{quote}



\begin{quote}

P4: \textit{``
I was afraid that if there were one or two people, they themselves would be recognized and appear, but if there were more, they would become a single group of “several people. The fear of not only how things themselves are transformed, but also how “what becomes one thing” is determined.
''}
\end{quote}

\begin{quote}

P6: \textit{``
I am afraid. I don't like reality being rewritten. I dislike how it takes the liberty to take what I am seeing and put together what the key points are without my permission and interpret them without my permission. I also felt that interpretation intervenes very much when generating images from text.
''}
\end{quote}

Similarly, some participants recognized the dizzying and inconsistent changes in background and other areas that were not considered important as a characteristic of AI, but this may be a result of the reliance on this implementation, which did not take temporal connections into account.
Many participants associated this lack of consistency in the background with dreams.

\begin{quote}

P5: \textit{``
The lack of background and contextual context is AI's shortcoming, and that's what made it interesting.
''}
\end{quote}

\begin{quote}
P6: \textit{``
Dreamlike. The concept was kept, but the details were inconsistent and the jumps were close to the dream experience.
''}
\end{quote}



On the other hand, there were participants who did not fully immerse themselves, but enjoyed the act of reasoning about the sentences that were being passed along the way in their goggles. It is a reasoning that traces our algorithm backwards from common sense and prior knowledge about the room.

\begin{quote}

P7: \textit{``
I was looking at the view, trying to figure out what kind of words were being used.
''}
\end{quote}

\begin{quote}
P6: \textit{``
I enjoyed the feeling of indirect reasoning. To think about the landscape that brought that sentence down from what you actually have to the sentence. It's in the picture of what you actually have and the picture you're in, different but in the same category and so on, so the process of deducing the difference between the two is interesting.
''}
\end{quote}

\vskip\baselineskip

\textbf{Subjective understanding of semantic/linguistic mediation}

Gradually, participants begin to recognize these issues and sensations, which they initially associated with the technology of AI, as heavily dependent on the very structure of communicating views via linguistic expressions.

\begin{quote}

P7: \textit{``
At first I thought it was a subjective experience of the AI process, but then I began to realize that this is no different than what happens when communicating verbally. You mentioned that you were cutting down on the information in front of you, and I had the feeling that I was doing exactly that when I was communicating with others. When I communicate my own experiences, I am reducing them to short sentences, so I feel that what I am doing is the same.
''}
\end{quote}

\begin{quote}
P5: \textit{``
I think the result would be much the same whether it was done by a person or an AI. The final texturing and such are different, though. Especially the texturing might be almost the same.
''}
\end{quote}

\begin{quote}
P9: \textit{``
They were all Westerners. I think it is because of the assumptions of this model. This appears to be an AI problem, but it is a language problem. When describing a scene domestically, you wouldn't dare use the words “a Western man.”
In verbal communication, it is important that not only what is important is verbalized, but also that prerequisite knowledge and other information is omitted.
''}
\end{quote}

\begin{quote}

P9: \textit{``
It is similar to the feeling you get when you are doing a sika-wari game (watermelon splitting).
People around you say “right” and “left,” and the textual information is converted in your head and...
''}
\end{quote}

Some participants noticed this not during the experience, but by looking at the outside display of others who were experiencing it.

\begin{quote}
P2: \textit{``
While at first I attributed the unpleasantness entirely to characteristics on the part of the AI, when I saw the characters mediated by others in their experience, I realized that it was also a characteristic of the verbalization itself.
''}
\end{quote}

Some participants also compared this to the fact that not only communication with others through language, but also their own perceptions and memories are already linguistically mediated. When sensory information is mediated from the past to the future through memory, there will be a process similar to that of linguistic mediation with others.

\begin{quote}
P5: \textit{``
Visual memories of being small can feel as if they disappear when put into writing. That is how we make memories smaller.
''}
\end{quote}

Furthermore, participants were. thus puzzled by the complexity of the issue, which they understood as universal to the mediation of sensory information by language.

\begin{quote}
P6: \textit{``
I was afraid to average it out and interpret it on my own. Both image to text and text to image. But what should I do if I didn't average? I'm not sure when I think about it.
''}
\end{quote}

\begin{quote}

P2: \textit{``
It seemed to me that it was more of a daunting problem to see it as a general language problem than to see it as an AI problem. I thought it could be solved to some extent by eliminating model biases, etc. and adapting it to human society.
''}
\end{quote}

\section{Discussion}

This section discusses the device of Semantic See-through Goggles and the conceptual possibilities it suggests. First, we describe the technical limitations and possible implementation of the current prototype. Then, we explain the features and advantages of Semantic See-through Goggles as a methodology for understanding linguistic mediation by AI. Finally, we propose the concept of ``Linguistic/Semantic Virtual Reality'' as suggested by Semantic See-through Goggles.

\subsection{Limitation and Possibility of Implementation}
Here we describe the prototype limitation and envisioned variations in order to discuss the implementation possibilities of the Semantic See-through framework.

\paragraph{Precision and time performance}
In general, there is a trade-off between model accuracy and inference speed (time performance). AI models, such as high-precision image captioning and image generation models, typically require more computational resources and processing time, meaning that improving accuracy often leads to increased inference time\cite{ang2022characterizing}. 
In the prototype, the number of characters in the image captioning process and the inference steps in the image generation process were adjusted to balance accuracy and inference time. However, it should be noted that these parameters can affect the semantic aspects of the output. The number of characters controls the level of detail in the image description, while the inference steps determine the level of detail in the visual representation. In other words, these parameters define the boundary between the preserved meaning and the meaning that is lost. In this system, the number of characters is set to 20-50 for convenience, and the number of inference steps is set to 4. However, it is necessary to further investigate how tuning these parameters affects the output, as changes in these settings may significantly alter the results.
In the future, considering that the human visual system is relatively tolerant of delays up to approximately 100 milliseconds\cite{10.1145/634067.634255}, the performance of the Semantic See-Through experience can be further improved by selecting a highly accurate model while maintaining processing latency within this limit.

\paragraph{Personhood}
Sentences in typical image captioning are narrated in the third person. It is because most image captioning models aim to capture overall content to balance readability, information volume, and avoid information overload\cite{ng2020understanding}. For example, when describing a person touching an apple in front of them, an image-captioning model would generate the sentence 'A person is touching an apple.' However, for the user wearing the goggles, 'I am touching an apple' would be more appropriate. While the user's perspective is first-person, the captions are generated in the third person. This also affects image generation, leading to a continuous output of images where the viewpoint is inconsistent. By incorporating external information, such as converting 'A person' to 'I,' it may be possible to generate a more subjective perspective.
\paragraph{Spatial information}
In prototype image captioning, the text often lacks information about spatial placement because it focuses on describing the overall content. It is because the Transformer architecture, which is currently the mainstream, does not recognize positional information. This is because spatial information is lost when visual features are flattened and converted into a sequence format. As a result, in prototypes, the spatial arrangement of objects is not reflected in the text, leading to discrepancies between the arrangement of objects in the output image and the original image. However, this issue may be mitigated by adding external, subjective spatial position information. For example, describing “An apple” as “An apple on the left” may result in an output that better preserves spatial information. In the future, it might be possible to achieve outputs that retain spatial arrangement by using an image-captioning model that takes a 3D spatial information as input\cite{Ahmed2023Enhancing}, or a generative model that generates a 3D scene from text\cite{zhang2024text2nerf}.
\paragraph{Temporal information}
In prototype image captioning, the input is treated as a single image, meaning that each image is processed independently, and the temporal relationships between images are lost. It is becasuse when an entire image is compressed into a single static feature in a standard model, the temporal correlation and motion information between successive frames in a video are lost\cite{7984828}. As a result, in prototypes, what should be described as “A person is putting down an apple,” for example, might be incorrectly captioned as “A person is picking up an apple” because the movement of the object is not captured. However, it may be possible to better capture the meaning of the motion by externally adding temporal information, such as converting “an apple” into “an apple which a girl tossed to me.” In the future, outputs that retain temporal information could be achieved by using an image-captioning model that takes video as input\cite{Gao2017Video} or a generative model that generates video from text\cite{Ho2022Imagen}.

\paragraph{Different level of granularity}

In prototype image captioning, objects in the image are often described with general knowledge, leading to a loss of input specificity. It is because image captioning is typically designed to describe images at a fixed level of granularity, as descriptions tend to be anchored to broader categories or concepts\cite{wang-etal-2014-poodle}. 
However, humans use different granularities for different situations, animals, dogs, golden retrievers, etc. In the future, when this point is resolved on the part of the model, this choice of granularity will also reflect the bias of the model, which will be reflected in the experience of Semantic See-through Goggles.


\subsection{SSTG for understanding AI cognitive process}
The proposed system compresses information into a single line of text once the view is in the process of reaching the display image from the camera image. This allows participants to subjectively experience what is lost in the representation of the world linguistically understood by AI and what biases are included to compensate for what is lost. In fact, the workshop and its analysis allowed participants to experience, sometimes with surprise and conviction, the various characteristics associated with AI and its problems. Here, we reiterate the unique advantages that this frame has as a tool for overviewing the perception of AI.

\subsubsection{Subjectivizing the problem}
At first glance, the nature of the AI noticed by the Semantic See-through Goggles experience seems to be a point that has already been pointed out by many researchers (even if they themselves did not know it), especially as participants commented that object abstraction and averaging depend on the bias of the data. However, what they experienced as they observed the world in which their bodies existed and acted on the basis of their observations brought about a strange sensation and brought physical and emotional understanding. This tool can help those who lack knowledge about the seriousness of AI and AI-related issues to understand them in a short time and with a strong impression. It is also suggested that it provides an opportunity for AI researchers to rethink the problems they are tackling on a daily basis by mobilizing their own physical senses. However, as mentioned at the beginning of this paper, we concentrated on the implementation of image captioning and verbal image generation, which is clearly via verbal information. However, there are other implementations of AI mediated by sensory information that only potentially involve semantic information. Therefore, it should be noted that it is an oversimplification to call this experience “the experience of seeing the world as an AI in general.


\subsubsection{Generalizing problems as a matter of semantic/linguistic mediation}
As the participants themselves realized during the experience, the issues that are understood and experienced through this tool are also related to the issue of scenery being mediated through language in general. For example, suppose you tell your friend that you met a couple at a café yesterday. The listener would probably think of a couple of standard build, of the majority race and sexuality in the scene. This very act of verbal mediation of a scene has inherent problems of the same structure as image captioning and image generation by AI. When visual information is converted into a text, the environment is first isolated and selected under the bias of the subject or the community to which it belongs, and when the recipient reconstructs the scene, the bias of the subject and the community also causes the scene to be often averaged out and beautified. The proposed method allows participants to experience this misalignment as their own view in real-time and in first-person, and as a result, it was both surprising and discomforting to them in this workshop. As the participants also pointed out, this mediation occurs not only among others, but also in memory recall, where a similar phenomenon occurs as a linguistic mediation of the view from oneself to oneself. By simultaneously experiencing such a problem that we, as linguistic animals, cannot escape and the problem of linguistic mediation by AI in the form of a see-through HMD, the proposed method makes us recognize the problem of AI not as a transient technical problem, but as an inescapable problem for intelligence that sees the world only in terms of meaning.


\subsection{Linguistic Virtual Reality}
Virtual Reality always involves the question of ``what is the Virtuality of Reality''.
The general answer is that it reproduces the stimuli it gives to the sensory organs. For example, in sight, the identity of the image projected on the retina is virtuality; in hearing, it is the identity of the vibration of the eardrums, which is virtuality. In Sutherland, a mechanism that could arbitrarily control the molecules of surrounding objects could be considered the ultimate display, in the sense that it could reproduce the same stimuli as reality in any situation for any sensory organ. It can be a wall-type display like CAVE, a display in front of the eyeball, or, in recent years, a machine that projects directly onto the retina with a laser.

We will put as one answer to this question the identity of ``the sentence when we put it into words.'' When we see or hear something, we can put that perceptual experience into words. We can then transmit those words to others and to ourselves, again imaginatively reproducing the perceptual experience. Here, there are multiple pieces of perceptual information that are dropped, i.e., the same sentence just as there are multiple solids that realize the same retinal image by losing information when a certain solid is mapped onto the retina.

\begin{figure}[thb]
 \centering 
 \centering 
 \includegraphics[width=90mm]{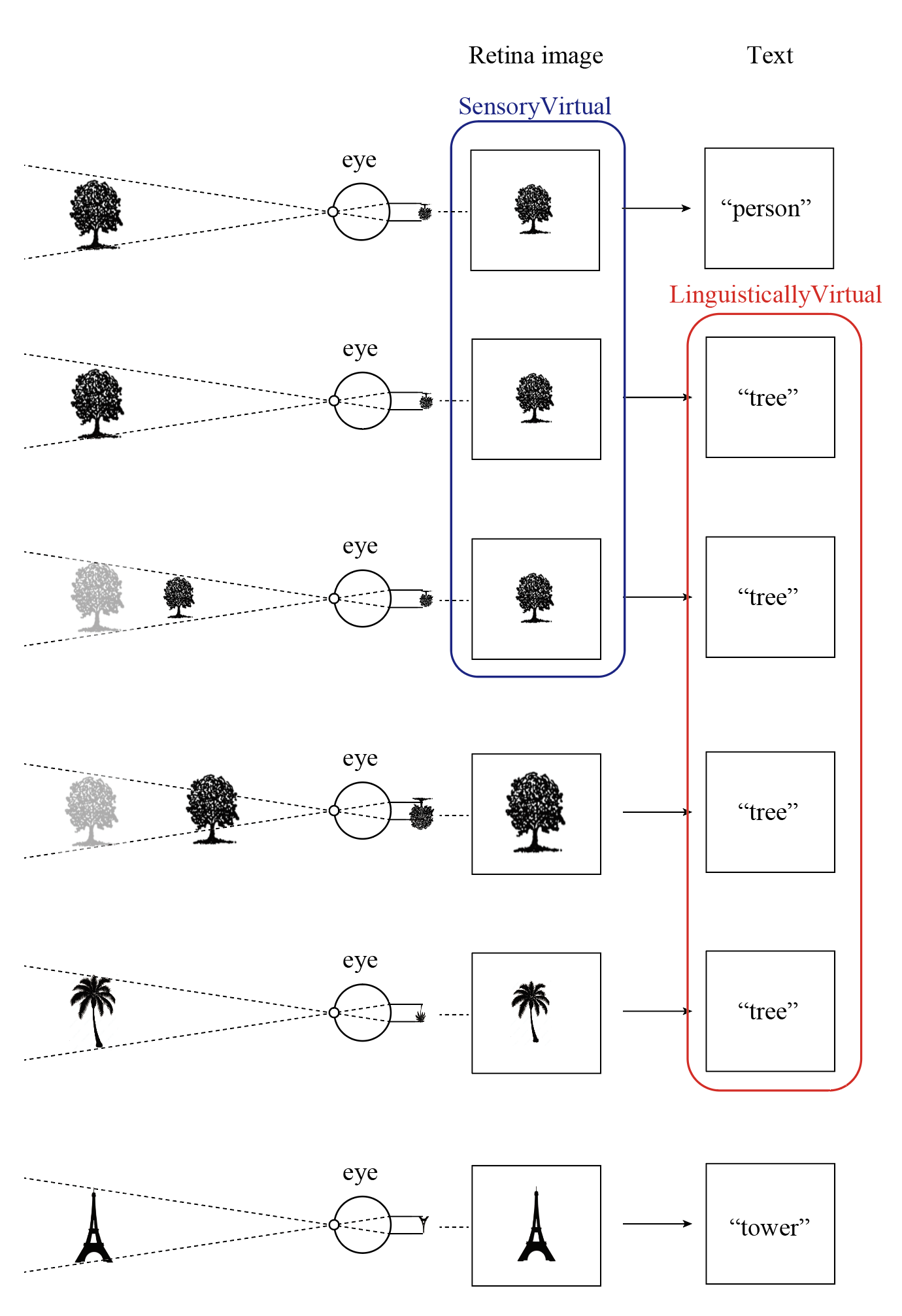}
 \caption{Linguistic/Semantic Virtual Reality.}
 \label{fig:svr}
\end{figure}

Figure \ref{fig:svr} sketches a comparison between sensory virtual reality in vision (investment diagram method and HMD-like virtual reality) and linguistic virtual reality. (1)-(3) are sensory equivalents of each other, since the image on the retina does not change. However, (4) is not equivalent because the image on the retina is larger, and (5) and (6) are not equivalent because not only the scale but also the visual image changes significantly. In linguistic virtual reality, however, we consider the equivalence when the visual image is further transformed into language. When the visual image becomes language, it can be thought of as when it becomes a sentence in the mind, or when it is conveyed to another person as a sentence, etc. Here, (2)-(5) are equivalent to each other because they are “trees” in terms of words, although they are different in size and type. On the other hand, (6) is not equivalent because it is a “tower. Furthermore, it is important to note that (1) and (2) form the same retinal image, but (1) is not linguistically equivalent because it is mistaken for a person due to some influence, and thus becomes a “person. Thus, the relationship between sensory virtual reality and linguistic virtual reality is not a simple one in which the latter is less virtual.

Our project demonstrates how this kind of linguistic equivalence can be superimposed on Sutherland's HMD, one symbol of VR in the usual sense, so that it can be seen as a virtual reality. We believe that this idea itself, as a conceptual device, has the potential to connect the sociological study of linguistic communication with the study of ethics in AI and the study of virtual reality.

\section{Conclusion}
In the paper, we proposed "Semantic See-through Goggles," an experimental framework for glasses through which the view is once fully verbalized and re-imaged, and introduced an implementation and the workshop using this.
Realized experience offers a methodology for a subjective understanding of what it’s like to share the task of perceiving the world with an AI. At the same time, this understanding was also recognized as the one around the problem of the mediation of scenery (sensory information) through language itself.
We further summarized the limitations and possibilities of this implementation and proposed a Semantic/Linguistic Virtual Reality, a kind of virtual reality that is inherent in the mediation of senses by language on intelligence: multiple realities that become the same sentence are equivalent. 
We hope that this research will connect the ethical issues of AI, the issue of reality/virtuality in VR and the issues of linguistic communication, appropriating some of the energy of the recent debate over AI for a classical exploration of the banal and inescapable fact that (natural or artificial) intelligence can only see the world under meaning. 


\bibliographystyle{ACM-Reference-Format}
\bibliography{sample-base}


\end{document}